\documentclass[conference]{IEEEtran}
\IEEEoverridecommandlockouts

\pdfoutput=1

\usepackage{amsmath}
\DeclareMathOperator*{\argmin}{argmin} 
\DeclareMathOperator*{\argmax}{argmax}

\usepackage[vlined,linesnumbered,ruled]{algorithm2e}
\usepackage{subcaption} 
\usepackage{tikz}
\newcommand*\circled[1]{\tikz[baseline=(char.base)]{
            \node[shape=circle,draw,inner sep=1pt] (char) {#1};}}
\usepackage{tabularx} 

\def\ie{\textit{i.e.}\xspace}

\def\etal{\textit{et al.}\xspace}

\def\eg{\textit{e.g.}\xspace}

\usepackage{cite}
\usepackage{amsmath,amssymb,amsfonts}
\usepackage{algorithmic}
\usepackage{graphicx}
\usepackage{textcomp}
\usepackage{xcolor}
\def\BibTeX{{\rm B\kern-.05em{\sc i\kern-.025em b}\kern-.08em
    T\kern-.1667em\lower.7ex\hbox{E}\kern-.125emX}}
\begin{document}

\title{ Accelerating End-Cloud Collaborative Inference via Near Bubble-free Pipeline Optimization}

\author{\IEEEauthorblockN{Luyao Gao$^{1,2}$, Jianchun Liu$^{1,2}$, \ Hongli Xu$^{1,2}$, Sun Xu$^{1,2}$, \ Qianpiao Ma$^{3}$, \ Liusheng Huang$^{1,2}$   \\}
\IEEEauthorblockA{  
$^1$School of Computer Science and Technology, University of Science and Technology of China, China\\
$^2$Suzhou Institute for Advanced Research, University of Science and Technology of China, China\\ 
$^3$ School of Computer Science and Engineering, Nanjing University of Science and Technology, China\\
}

}

\maketitle

\begin{abstract}
End-cloud collaboration offers a promising strategy to enhance the Quality of Service (QoS) in DNN inference by offloading portions of the inference workload from end devices to cloud servers. 
Despite the potential, the complex model architectures and dynamic network conditions will introduce numerous bubbles (\ie, idle waiting time) in pipeline execution, resulting in inefficient resource utilization and degraded QoS.
To address these challenges, we introduce a novel framework named COACH, designed for near bubble-free pipeline collaborative inference, thereby achieving low inference latency and high system throughput.
Initially, COACH employs an \textit{offline} component that utilizes an efficient recursive divide-and-conquer algorithm to optimize both model partitioning and transmission quantization, aiming to minimize the occurrence of pipeline bubbles.
Subsequently, the \textit{online} component in COACH employs an adaptive quantization adjustment and a context-aware caching strategy to further stabilize pipeline execution. 
Specifically, COACH analyzes the correlation between intermediate data and label semantic centers in the cache, along with its influence on the quantization adjustment, thereby effectively accommodating network fluctuations.
Our experiments demonstrate the efficacy of COACH in reducing inference latency and enhancing system throughput.
Notably, while maintaining comparable accuracy, COACH achieves up to 1.7$\times$ faster inference and 2.1$\times$ higher system throughput than baselines.

\end{abstract} 


\section{Introduction}\label{sec:intro}
Traditionally, the DNN-based Artificial Intelligence (AI) applications (\eg, model inference tasks) are usually offloaded to powerful cloud servers for advanced processing, facilitating faster and more accurate DNN inferences \cite{dillon2010cloud}. 
However, transmitting data to a server raises privacy concerns, particularly with sensitive information.
Additionally, many resource-limited end devices struggle with the demands of complex DNN tasks, such as real-time high-definition video processing, which underscores a capability gap for less powerful devices \cite{xu2022adaptive}.
In order to address the disparity in computing resources, collaborative inference involving both end devices and cloud servers has emerged, enabling DNNs to be segmented for distributed processing \cite{hu2019dynamic, almeida2022dyno, yang2022cnnpc}. 
The process initiates with the end device executing the initial DNN segment and sending the intermediate data to the server, which then completes the processing and returns the final result \cite{kang2017neurosurgeon}. 

Current approaches of collaborative inference always aim at the following two goals.
(1) \textbf{Low inference latency.}
Quick decision-making is crucial in time-sensitive applications such as autonomous vehicles, which necessitate the swift processing of sensor data within 20ms \cite{sonko2024comprehensive}.  
To facilitate this, DNN models are typically partitioned into two segments for parallel processing, with task latency comprising three parts: end device computation latency, intermediate data transmission latency, and server computation latency \cite{kang2017neurosurgeon}.
To achieve low latency, it is essential to explore efficient DNN partitioning and compression strategies that take into account the diverse capabilities of end devices and servers \cite{li2018jalad, liu2023finch}.
(2) \textbf{High system throughput.} 
In practice, end devices often need to process continuous inference tasks, necessitating efficient scheduling strategies to enhance the throughput of inference systems \cite{hu2019dynamic}.
Meanwhile, due to complex models and high workloads, the latency of the cloud computation stage cannot be ignored and should also be considered in scheduling \cite{dillon2010cloud}. 
To this end, pipeline parallelism is a crucial technique, allowing for the overlap of computation and transmission stages \cite{qi2024zero}.
However, inference efficiency will be compromised by numerous pipeline bubbles (\ie, idle waiting time), caused by unbalanced stage execution times \cite{duan2023optimizing}.
Thus, reducing bubbles during pipeline execution is crucial for enhancing the inference performance.


However, two major challenges related to model partition and transmission extremely hinder the reduction of bubbles. 
(1) \textbf{Complex model architecture.} 
The evolution of DNN designs from linear to more sophisticated structures, such as Directed Acyclic Graphs (DAGs) exemplified by GoogleNet \cite{simonyan2014very} and ResNet \cite{he2016deep}.
The multiple dependencies in DAG models create exponential search spaces for optimal partitioning and compression strategies, complicating pipeline parallelism \cite{almeida2022dyno}.
Moreover, the intricate connections within these models often introduce more bubbles in the pipeline, further dragging the inference process \cite{yao2022edge,liao2023accelerating}.
(2) \textbf{Dynamic network conditions.}
The effectiveness of collaborative inference is substantially influenced by network conditions, which are inherently variable in real-world applications \cite{li2018edge}.
Although carefully crafted scheduling strategies can minimize latency in the pipeline execution, numerous bubbles still arise due to dynamic networks, resulting in high transmission latency and reduced system throughput \cite{ren2022survey}. 

Existing collaborative inference approaches primarily aim at reducing task latency and improving system throughput from the following two aspects. 
First, some works \cite{banitalebi2021auto, yang2022cnnpc, almeida2022dyno} focused on optimizing the DNN partition strategy for complex models, yet they inadequately tackled the issues related to pipeline parallelism across varied computing resources of both end devices and server. 
Second, other works \cite{li2021throughput, duan2021computation, duan2023optimizing} concentrated on parallel processing to manage the multiple dependencies and complex interactions in DAG models, but they generally failed to account for the emergence of bubbles in pipeline execution of continuous tasks. 
This oversight frequently leads to the accumulation of numerous bubbles in the pipeline.
Additionally, these approaches were typically designed for static conditions and experienced substantial performance degradation under dynamic networks \cite{yao2022edge}.


To address the aforementioned challenges, we present COACH, a novel collaborative inference framework designed for near bubble-free pipeline execution, including an \textit{offline} component and an \textit{online} component. 
COACH is engineered to achieve low latency and high throughput under dynamic network conditions by minimizing bubbles in the pipeline.
It incorporates an \textit{offline} component that layer-wisely determines the joint model partitioning and transmission quantization strategy, and carefully manages layer parallel execution.
Subsequently, the \textit{online} component facilitates real-time assessment of the status of task features, enabling an adaptive quantization adjustment and a context-aware caching strategy to stabilize the pipeline execution in dynamic networks.

Nevertheless, to achieve efficient collaborative inference performance, two principal challenges still need to be addressed in COACH. 
First, we observe that different layers within DNN models require varied levels of quantization precision to satisfy specific accuracy criteria \cite{almeida2022dyno}, as shown in Section \ref{observations}. 
This observation highlights the complex and tight interaction between partitioning and quantization strategies, further complicated by layer dependencies and pipeline parallelism.
Consequently, it is critical to \textit{determine the optimal partitioning and quantization strategy} to achieve bubble-free pipeline execution.
Second, if pipeline scheduling fails to adapt to network fluctuations, the increase in transmission latency may lead to substantial pipeline bubbles, underscoring the need for quantization adjustment to stabilize transmission.
Thus, another challenge involves how to \textit{implement adaptive quantization adjustment in dynamic networks} to enhance collaborative inference performance. 
The key contributions of this paper are as follows:
\begin{itemize}
    \item We present COACH, a novel collaborative inference framework designed for near bubble-free pipeline execution. 
    To the best of our knowledge, COACH is the first framework that integrates the \textit{offline} and \textit{online} components to optimize the collaborative inference scheduling by sufficiently minimizing pipeline bubbles. 

    \item The \textit{offline} component utilizes a novel recursive divide-and-conquer algorithm to jointly determine optimal model partitioning and transmission quantization strategies, thereby achieving balanced and efficient computation and transmission phases.
  
    \item The \textit{online} component implements an adaptive quantization adjustment and a context-aware caching strategy to accommodate network fluctuations.
    It adjusts the quantization precision according to the correlation between intermediate data and the semantic centers in the cache.

    \item Extensive experimental results demonstrate that COACH significantly surpasses existing collaborative inference approaches.
    Notably, while maintaining comparable accuracy, COACH achieves up to 2.1$\times$ faster inference and 2.5$\times$ higher throughput compared to baselines.
    
\end{itemize}

\section{Background and Motivation}\label{sec:motivation}
\subsection{Related Works}

The collaborative inference paradigm, which leverages strategic workload offloading, aims to expedite the DNN inference process across heterogeneous end devices and servers \cite{kang2017neurosurgeon, hu2019dynamic, almeida2022dyno, yang2022cnnpc}. 
Optimization techniques generally involve model partitioning, transmission compression, and dynamic scheduling.

A significant advancement in \textit{model partitioning} is provided by Neurosurgeon \cite{kang2017neurosurgeon}, which reduces latency through tailored partitioning strategies for chain topology models.
IONN \cite{jeong2018ionn} adopts the shortest path algorithm to enhance simultaneous inference processing, representing a novel methodological shift. 
OfpCNN \cite{yang2023ofpcnn} advances model partitioning with a fine-grained approach that optimally utilizes heterogeneous resources on devices and servers. 
Nonetheless, model partitioning introduces an increased transmission overhead, potentially complicating the partitioning search space \cite{eshratifar2019jointdnn, liu2023yoga}.
Recent advancements in DNN \textit{transmission compression} techniques, such as quantization, have been directed towards tackling the challenge of high transmission overhead, thereby reducing transmission requirements \cite{almeida2022dyno,yang2022cnnpc,li2018jalad,banitalebi2021auto}.
Approaches like CNNPC \cite{yang2022cnnpc} and auto-split \cite{banitalebi2021auto} employ quantization to enhance the potential for collaborative inference execution, though they may encounter challenges in adaptation across diverse end-cloud environments.

Additionally, focusing exclusively on optimizing a single task is inadequate for managing continuous task scenarios. 
It is essential to concurrently schedule multiple inference tasks for efficient execution, particularly when employing common practice pipeline scheduling \cite{ren2022survey, shi2019improving}.
The \textit{scheduling complexity} inherent in collaborative execution across devices and servers, particularly within DAG topology models, has emerged as a critical area of focus \cite{hu2019dynamic, huang2020clio, laskaridis2020spinn, duan2021computation}.
Hu \etal \cite{hu2019dynamic} and Li \etal \cite{li2021throughput} explore the complexities of optimal offloading scheduling for DAG models in continuous task scenarios, yet their approaches primarily focus on static environments and tend to increase the occurrence of bubbles in dynamic networks.
Duan \etal \cite{duan2021computation} and Zeng \etal \cite{zeng2020coedge} investigate DNN partitioning and pipeline parallelism strategy, focusing on throughput enhancement for DAG models.
However, their approaches primarily optimize pipeline scheduling for parallel execution, while neglecting the cloud computation stage and the latency tolerance of individual inference tasks.

\subsection{Key Observations in Inference Tasks}\label{observations}

\begin{figure}[t]
    \centering
    \begin{subfigure}[b]{0.45\textwidth}
    \includegraphics[width=\textwidth]{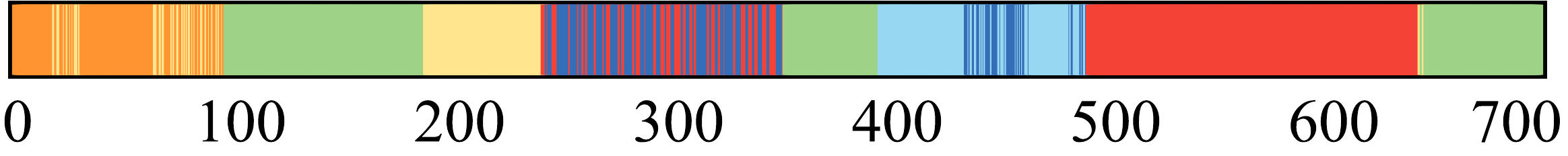}
    
    \caption{Time locality visualization as a color line, in which similar colorations denote similar features.}\vspace{1mm}
    \label{fig:time_cor}
    \end{subfigure}
    \hfill
    \vspace{1.5mm}
    \centering
    \begin{subfigure}[b]{0.35\textwidth}
    \centering
    \includegraphics[width=\textwidth,alt={correlation}]{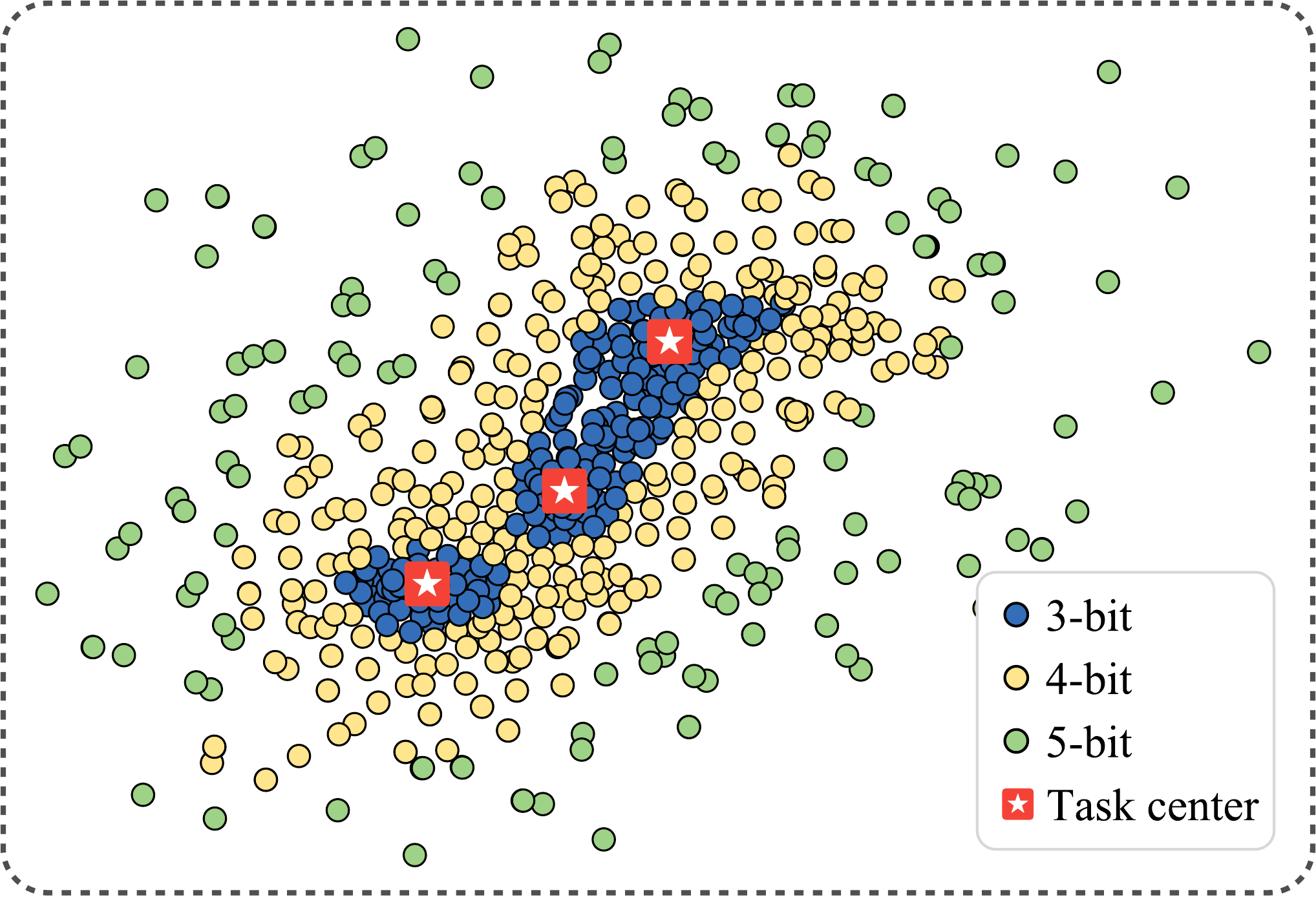}
    \caption{Spatial locality visualization as 2D points, color-coded by optimal quantization precision.}
    \label{fig:data_cor}
    \end{subfigure}
    \vspace{-1mm}
    \caption{Data correlation visualization on the UCF101 dataset with the ResNet101 model.}
    \label{fig:correlation}
    \vspace{-5mm}
\end{figure}

In collaborative inference tasks, like user interactions and video recognition in end devices, the data often exhibits repetitive patterns \cite{li2021boosting}. 
These patterns lead to the generation of similar intermediate data during the inference process, providing a solid foundation for predictive optimization in subsequent tasks. 
By leveraging the predictable nature of data correlations, substantial benefits can be achieved, including streamlined computations and efficient pipeline execution \cite{shi2016edge}. 
To highlight the importance of data correlation, we focus on both the temporal and spatial localities of intermediate data using the ResNet101 model \cite{liu2024federated} on the widely-used UCF101 video dataset \cite{soomro2012ucf101}. 
Additionally, building on experiences in Section \ref{sec:pfmc} with the ImageNet-100 dataset \cite{krizhevsky2012imagenet}, we demonstrate the versatility of our observations across various scenarios.
We analyze the intermediate data across frames over extended periods, utilizing the t-SNE technique \cite{van2008visualizing} to visualize the data features as a color line. 
Fig.~\ref{fig:correlation}(\subref{fig:time_cor}) demonstrates pronounced consistency in the coloration of features over short intervals, indicating strong temporal locality among the intermediate data.
This observed correlation supports the feasibility of temporarily caching previously computed results to accelerate future collaborative inference processes.

In Fig.~\ref{fig:correlation}(\subref{fig:data_cor}), the data points are color-coded according to their optimal quantization precision (3-bit, 4-bit, or 5-bit), illustrating distinct clusters of inference tasks. 
Notably, data points with 3-bit precision typically cluster closer to the task center, whereas those with 5-bit precision are more widely dispersed. 
This observation indicates that data points more resistant to clustering necessitate higher quantization precision to preserve inference accuracy. 
The analysis of spatial locality reveals a clear correlation between task specificity and quantization efficiency, emphasizing the potential to enhance inference processing by customizing quantization precision according to the variability of task-specific features \cite{li2021boosting}. 
The predictable nature of data correlations, as demonstrated in Fig.~\ref{fig:correlation}, supports the implementation of an efficient early-exit policy and adaptive quantization adjustment. 
The above strategies enhance transmission efficiencies, contributing to bubble-free pipeline scheduling from a novel perspective. 

\subsection{Potential Opportunities of Collaborative Inference} 
\begin{figure}[t]
    \centering    
    \includegraphics[width=0.41\textwidth,alt={workflow1}]{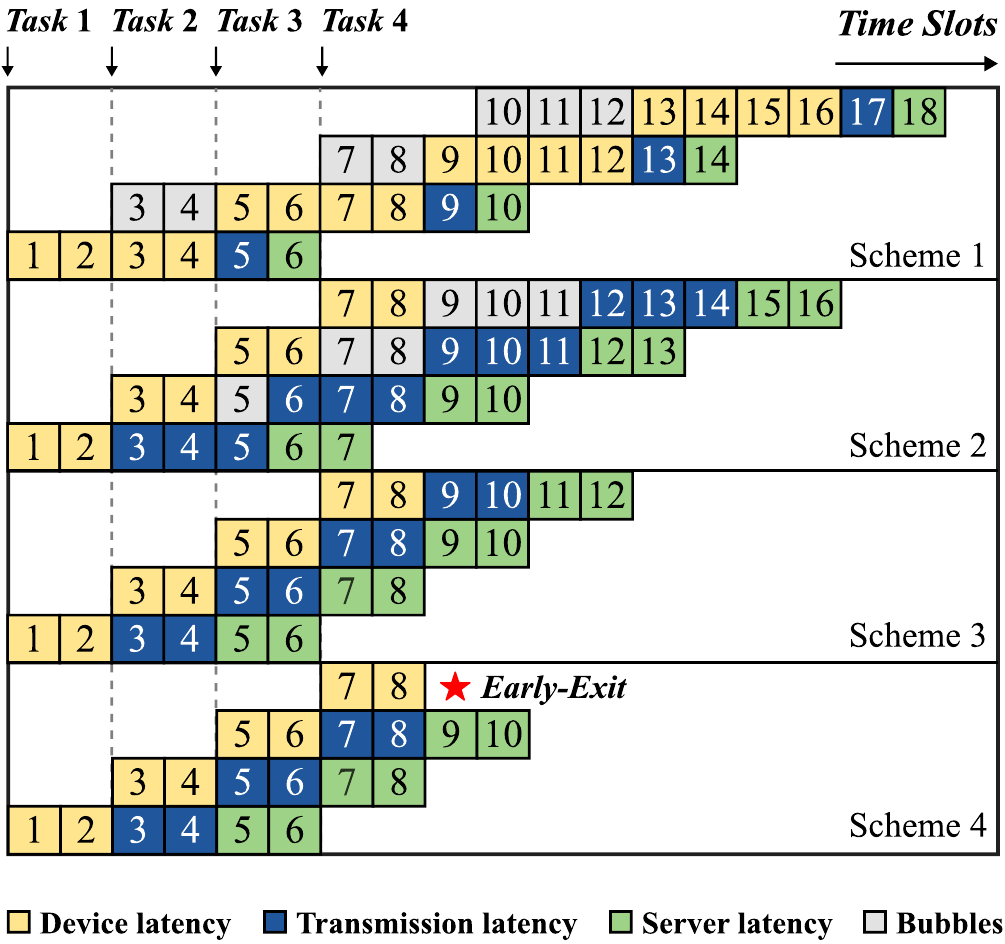}
    \caption{Three-stage collaborative inference processes with  pipeline scheduling.}
    \label{fig:workflow1}
    \vspace{-5mm}
\end{figure}

The essence of collaborative inference lies in optimizing the utilization of heterogeneous computing resources across end devices and servers. 
Fig.~\ref{fig:workflow1} illustrates the three-stage collaborative inference processes, where four inference tasks arrive sequentially every 2 time units. 
Scheme 1 is designed to minimize latency per task, achieving a duration of 6 time units per task \cite{bajic2021collaborative}.
Though Scheme 1 employs pipeline parallelism, it does not fully optimize resource utilization, leading to numerous bubbles in pipeline execution.
In contrast, Scheme 2 employs an alternative partitioning strategy specifically aimed at reducing bubbles in pipeline scheduling, which slightly increases task latency to 7 time units but significantly boosts system throughput.
Scheme 2 effectively manages both transmission and computation stages, resulting in fewer bubbles and enhanced system efficiency \cite{qi2024zero}.
This strategy demonstrates a practical advantage, achieving a 25\% efficiency increase over Scheme 1, as the maximum stage is reduced from 4 to 3 time units in pipeline execution.

Building upon the strengths of Scheme 2, Scheme 3 introduces an adaptive quantization adjustment that capitalizes on intrinsic data correlations and task-specific features to further minimize pipeline bubbles. 
This enhancement, as explored in Section~\ref{observations}, moves to achieve bubble-free pipeline execution and results in a reduction of 50\% compared to Scheme 1, as the maximum stage is reduced from 4 to 2 time units.
Furthermore, Scheme 4 integrates an early-exit policy that capitalizes on the temporal localities of inference tasks to streamline the inference processing.

In summary, as demonstrated in Fig.~\ref{fig:workflow1}, these strategies in Schemes 2-4 show significant potential in minimizing pipeline bubbles during the inference process. 
The following sections will further explore meticulously designed strategies 
with offline and online components, aimed at realizing near bubble-free pipeline execution and enhancing collaborative inference performance in terms of latency and throughput.

\section{System Design of COACH}\label{sec:system} 
\subsection{System Overview}

To enhance the effectiveness of collaborative inference, we concentrate on refining the collaborative inference process and exploiting data correlations.
We introduce COACH, a near bubble-free collaborative inference framework to handle the complexities of DNN models and dynamic networks, ensuring efficient resource utilization while maintaining high accuracy. 
With the combined strengths of the offline and online components, COACH concurrently manages computation and transmission stages, preserving a fluid pipeline flow and minimizing bubbles in pipeline scheduling.
%

\begin{figure}[t]
    \centering
    \includegraphics[width=0.42\textwidth]{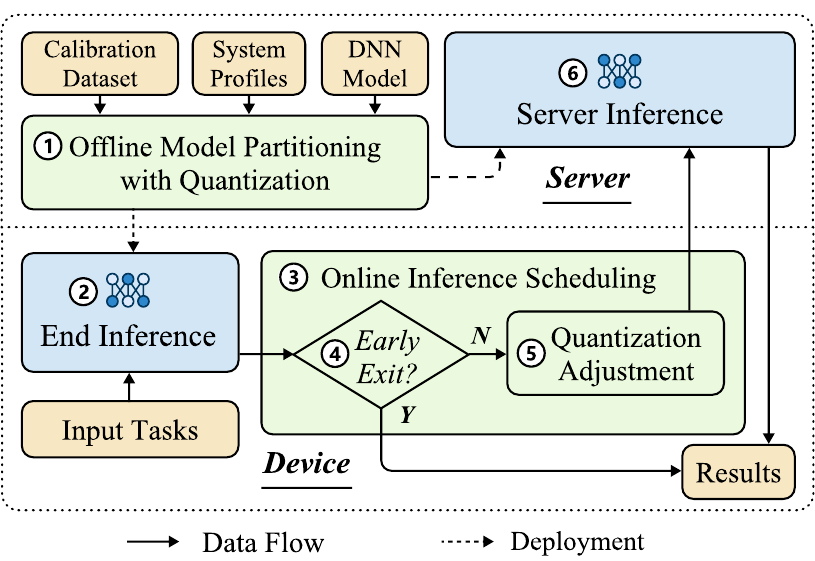}
    \vspace{-2mm}
    \caption{Overview and inference workflow of COACH.} 
    \label{fig:workflow}
    \vspace{-5.4mm}
\end{figure}

In Fig.~\ref{fig:workflow}, COACH starts with an offline component that handles model partitioning and quantization strategy (\circled{1}), acquiring system profiles and a calibration dataset.  
This component, which executes one-time before the inference process, carefully manages layers of parallel execution and initializes the online component without affecting inference performance. 
After that, the DNN model is deployed on both the end device and cloud server, setting the stage for collaborative inference execution with minimized pipeline bubbles.
During the inference process, input tasks are initially processed through the partitioned model on the end device (\circled{2}), generating intermediate data. 
Subsequently, the online scheduling component (\circled{3}) evaluates the task-specific quantization requirements of the intermediate data, facilitating adaptive quantization adjustment to stabilize pipeline execution in dynamic networks.
Once the quantization adjustment meets the early-exit condition (\eg, exceeding the early-exit threshold), the procedure will promptly generate the result (\circled{4}). 
Otherwise, COACH adjusts the quantization precision to further minimize pipeline bubbles before transferring the data to the cloud server (\circled{5}).
The cloud server conducts the remaining computations and returns the inference results to the end device (\circled{6}), thereby completing the inference task.

\subsection{Offline Model Partitioning and Quantization Component}
\label{subs:alg1}

We develop a recursive divide-and-conquer algorithm tailored for optimizing model partitioning and determining the quantization precision, as presented in Algorithm~\ref{alg:alg1}.

\textbf{Problem Formulation.}
To optimize transmission efficiency, COACH utilizes the Uniform Affine Quantization (UAQ) technique \cite{krishnamoorthi2018quantizing} to compress intermediate data efficiently while preserving accuracy. 
This quantization facilitates streamlined data transmission, allowing the cloud server to subsequently dequantize the data and continue the inference process. 
Optimal quantization precision is determined through a dichotomous search \cite{tsvetkov2014dichotomous,yan2024peaches}, based on the correlation between higher precision and improved inference accuracy.
We define $Q(v_i)$ as the quantization approach for layer $v_i$ in the DNN model, correlating with inference accuracy $Acc(\cdot)$ and constrained by accuracy loss limit of $\epsilon = 0.5\%$, thereby ensuring robust inference performance \cite{almeida2022dyno}:
\begin{equation}\label{eq11}
|Acc(v_i) - Acc(Q(v_i))| \leq \epsilon
\end{equation}

Let $V_e$ and $V_c$ denote the set of layers executed on the end device and cloud server, respectively, with $V_p$ representing the partition layer set. 
The latency of three stages in the pipeline can be defined as:
\begin{equation}\label{eq12}
    T_e = \sum_{v_i\in V_e} t_i^e, \quad T_t = \sum_{v_i\in V_p} Q(t_i^t), \quad T_c = \sum_{v_i\in V_c} t_i^c,
\end{equation}
where $t_i^e$, $Q(t_i^t)$, and $t_i^c$ represent the computation time on the end device, the transmission time with quantization, and the computation time on the cloud for layer $v_i$, respectively.
The task latency of collaborative inference is guaranteed in pipeline inference with $T_{\max}$ as:
\begin{equation}\label{eq13}
    T_e + T_t + T_c \leq T_{max}.
\end{equation}

\begin{figure}[t]
\centering
\includegraphics[width=0.45\textwidth]{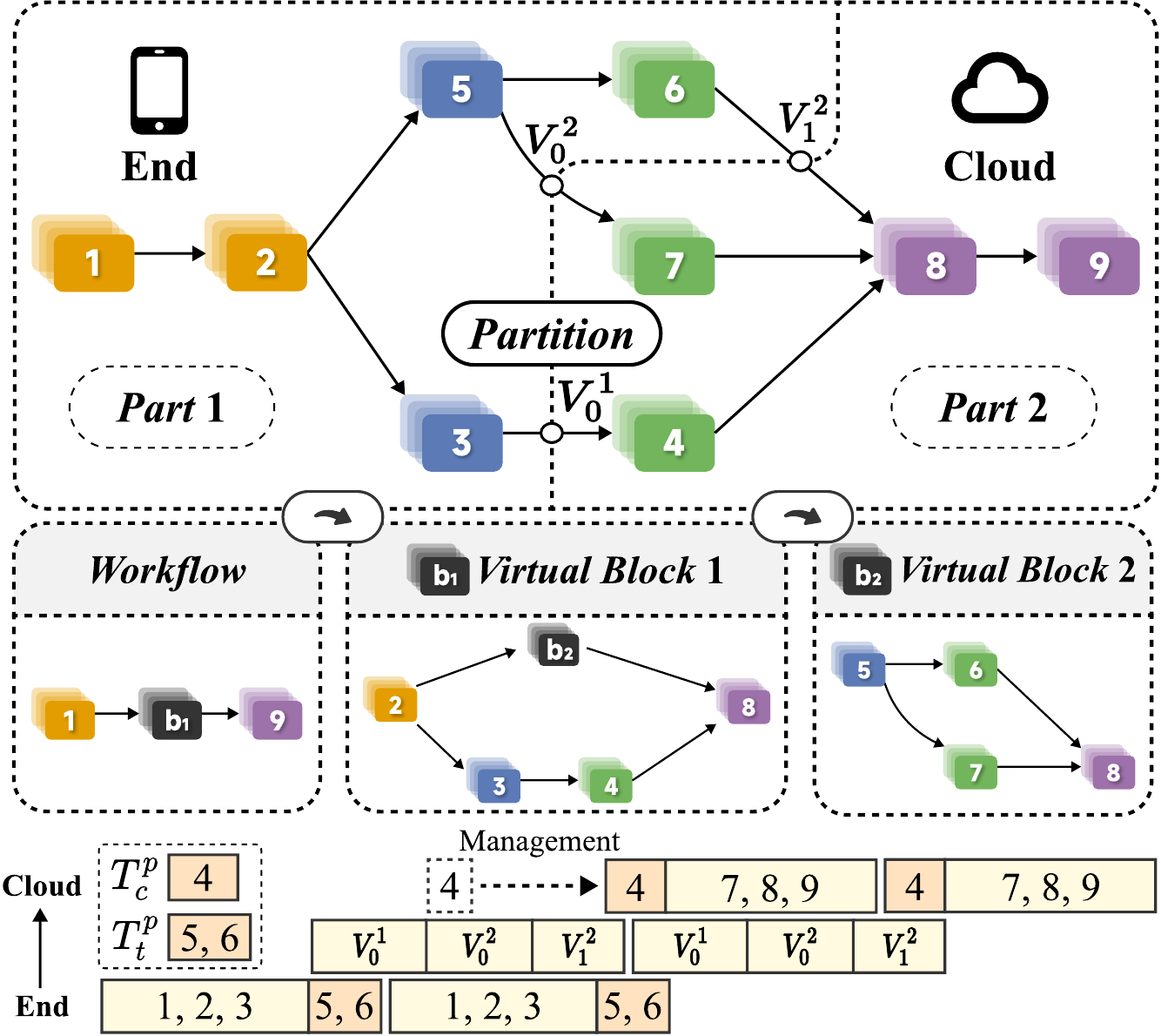}
\caption{Illustrating DNN partitioning with virtual blocks and layer parallel execution.} 
\label{fig:partition}\vspace{-5mm}
\end{figure}

Utilizing the DAG topology, layer parallel execution enables the simultaneous processing of non-interdependent layers.
As shown in Fig.~\ref{fig:partition}, after end device computing of layer 3, subsequent end device computing (\eg, layers 5 and 6) and transmission $V_0^1$ can execute in parallel.
After completing transmission $V_0^1$, the transmission $V_0^2$ and cloud computing of layer 4 can execute in parallel, while cloud computing of layer 7 waits for transmission completion, introducing additional pipeline bubbles.
By analyzing the layer dependencies, we carefully manage the layer parallel execution, considering the transmission and cloud parallel times $T_t^p$ and $T_c^p$, respectively, with the following constraint:
\begin{equation}\label{eq14}
T_t^p+T_c^p \leq \max\{T_e, T_t, T_c\}.
\end{equation}

Furthermore, we introduce two bubble functions for pipeline scheduling evaluation: $B_c(V_p)$ for computation bubbles and $B_t(V_p)$ for transmission bubbles, defined as:
\begin{equation} \label{eq15} 
    \begin{cases}
        B_c(V_p) &= \left|T_e-T_c\right|, \\
        B_t(V_p) &= \left|T_t - \max\{T_e, T_t - T_t^p, T_c - T_c^p\}\right|.
    \end{cases}
\end{equation}

The objective of the offline component is to identify an optimal partitioning and quantization strategy $V^*$ that maximizes the overall pipeline efficiency by minimizing the key bottlenecks, including bubbles and the maximum latency. 
The optimization problem for near bubble-free pipeline in COACH is formulated as follows:
\begin{equation} \label{eq16} 
\begin{aligned}
     V^* = \argmin_{V_p} &\left\{ B_c(V_p) + B_t(V_p) + \max\{T_e, T_t, T_c\} \right\} \\
    &s.t. \quad\quad   \ \eqref{eq11},\ \eqref{eq13}, \ \eqref{eq14}.
\end{aligned}
\end{equation}

\begin{algorithm}[h]
\caption{Offline and Online Components in COACH}
\label{alg:alg1}
\textbf{Offline Component:}\\
Evaluate layer computation and transmission costs in model $G$ with system profiles $P$.\\ \label{line1}
Cluster parallel layers into virtual blocks within $G$.\\\label{line2}
Determine chain flow $B_g = \{b_1, b_2, \ldots, b_n\}$ in $G$. \\
Initialize search space $S=\{B_g\}$ and offline strategy $V_p=\emptyset$.\\ \label{line3}
\For {each chain flow $B$ in $S$}{ \label{line4}
    Initial offline strategy $V_B=\emptyset$ for $B$.\\
    \For {each block $b$ in $B$}{ 
        Determine quantization precision for block $b$.\\
        Calculate bubble functions $B_c(V_B)$, $B_t(V_B)$.\\
        Assess $V_B$ performance in pipeline execution.\\ \label{line5}
        Update the optimal offline strategy $V_B$ for chain flow $B$.\\ \label{line6}
        \If {$b$ is a virtual block}{ 
            Append additional chain flows from $b$ to $S$.\\
        }
    }
    Update $V_p$ with $V_B$.\\  \label{line7}
}
Deploy the optimal strategy $V^*$ to device and server.\\ 

\textbf{Online Component:}\\
Initialize semantic centers $\mathbf{T_c}$ by calibration dataset $D$.\\ \label{line8}
Evaluate early-exit threshold $S_{ext}$ and quantization thresholds $\mathbf{S}_{adj}$.\\ \label{line9}
\While {Receive task feature $F$}{ \label{line10}
    Get similarity degrees $\mathbf{T}$ and task separability $S$.\\ \label{line11}
    \If{$S>S_{ext}$}{ 
        Update $T^c_j \in \mathbf{T_c}$ by Eq. \eqref{eq21}.\\ 
        Get inference result $R$ from $\mathbf{T_c}$ by Eq. \eqref{eq24}.\\ 
        Continue.\\ \label{line12}
    }
    Get real-time network bandwidth and estimate quantization requirement $Q_r$ with $\mathbf{S}_{adj}$.\\ 
    Quantize $F$ with $Q_c$ by Eq. \eqref{eq25} \\  \label{line14}
}
\end{algorithm}

\textbf{Algorithm Description.}
The principal challenge arises from the impracticality of exploring all possible partitions of a DAG, due to the exponential increase in the potential model partitioning search space as the model complexity grows.
For chain topology, partition layer set $V_p$ only comprises a single layer, whereas for DAG topology, $V_p$ comprises multiple layers.
We introduce a novel recursive divide-and-conquer algorithm to address the complexities of partitioning DAG topology models.
As illustrated in Fig.~\ref{fig:partition}, our algorithm is achieved by conceptually clustering parallel layers into virtual blocks and shifting the optimization focus towards a simpler chain topology, enabling carefully managed layer parallel execution (\eg, layer 4) to ensure minimal pipeline bubbles.

In Algorithm~\ref{alg:alg1}, the offline component initiates with evaluating the computation and transmission costs of each layer in model $G$, which are pivotal for determining the latency of pipeline stages (Line \ref{line1}). 
To simplify the exploration, the algorithm divides the complex DAG topology into more manageable chain topology segments, thus initiating a search space $S$ for recursive analysis. 
COACH clusters parallel layers into virtual blocks within model $G$, organizing these blocks into a sequential chain flow $B$, as $\{1,b_1, 9\}$ in Fig.~\ref{fig:partition}, which is incorporated in the search space as $S = \{B\}$ (Line \ref{line2}).

The algorithm proceeds by iteratively evaluating each chain flow within the search space $S$, aiming to identify the corresponding optimal partitioning and quantization strategy $V_B$.
For each chain flow $B$ in search space $S$, the algorithm initially utilizes a dichotomous search to precisely determine the appropriate quantization precision for the inference data flow.
Subsequently, by recognizing the dependency relationships among layers, it facilitates the analysis of parallel execution times $T_t^p$ and $T_c^p$. 
This analysis enables the calculation of both the computation and transmission bubble functions $B_c(V_B)$ and $B_t(V_B)$, thereby the optimal strategy $V_B$ is assessed based on the pipeline performance, as $V_0^1$, $V_0^2$ and $V_1^2$ in Fig.~\ref{fig:partition} (Lines \ref{line4}-\ref{line5}). 
%
Within the chain flow $B$, the algorithm sequentially examines each virtual block, excluding those unsuitable for the strategy from further analysis. 
Conversely, suitable virtual blocks are recursively integrated into the search space $S$ for optimization, ensuring thorough and efficient exploration of potential partitions, as $b_1$ and $b_2$ in Fig.~\ref{fig:partition}.
Following the comprehensive evaluation of $B$ with layer parallel execution management, the offline strategy $V^*$ is updated with $V_B$ (Lines \ref{line6}-\ref{line7}). 
After thoroughly exploring the search space $F$, the algorithm finalizes the most effective partitioning and quantization strategy $V^*$, as the strategy $\{V_0^1, V_0^2, V_1^2\}$ in Fig.~\ref{fig:partition}. 
For a DAG model composed of $n$ parallel data flows, each containing $c$ independent layers, the conventional approach exhibits a time complexity of $O(c^n)$ \cite{xu2023ago}. 
Our algorithm achieves a substantially reduced time complexity of $O(cn)$, illustrating a significant improvement in efficiency for optimizing complex DNN models.


\subsection{Online Inference Scheduling Component}
\label{subs:alg2}

The offline model partitioning and quantization strategy might not sufficiently ensure bubble-free pipeline execution. 
Additionally, dynamic network conditions significantly impact data transmission, potentially introducing bubbles in pipeline scheduling \cite{mohammed2020distributed}. 
To address this, we introduce an online inference scheduling component that dynamically adjusts the quantization precision and implements an early-exit policy, thereby stabilizing pipeline execution in dynamic networks. 
Crucially, different tasks require varied quantization precision to maintain inference accuracy, facilitating further optimization of pipeline scheduling, as detailed in Section \ref{observations}.
To make efficient real-time decisions on quantization adjustments, we propose a context-aware caching strategy that maintains label semantic centers. 
This strategy allows for quantization adjustments to be tailored to specific tasks, as outlined in the online component of Algorithm~\ref{alg:alg1}.
  
\textbf{Label Semantic Centers with Caching Mechanism.} 
To efficiently manage large volumes of intermediate data, COACH utilizes the Global Average Pooling (GAP) function \cite{lin2013network}, which concentrates on the core characteristics of data.
Applied to intermediate data of dimensions $\left< C \times H \times W \right>$, the GAP function simplifies the data into a reduced dimension of $\left< C \right>$, where $C$ denotes the number of channels, and $H$ and $W$ represent the spatial dimensions. 
This process generates \textit{task features} $F$ for intermediate data, which are concise feature vectors encapsulating essential characteristics. 

Moreover, to make similarity-based decisions for quantization adjustment, the caching mechanism maintains a \textit{semantic center} for each label, denoted as $\mathbf{T_c} = \{T_j^c\}$, where label $\ j \in \{0,\ldots,n\}$. 
The offline strategy determines the quantization precision based on intermediate data after quantization, ensuring the accuracy of the inference process (\eg, 0.5\% accuracy loss).
Moreover, by analyzing the spatial locality of task features, we observe that intermediate data clustering around the label semantic centers requires less information transmission to maintain accurate inference, allowing for more aggressive quantization strategies.

Evaluating the correlation between intermediate data and the label semantic centers in the cache allows for dynamic adjustment of quantization precision, accommodating network fluctuations and minimizing pipeline bubbles. 
These label semantic centers are initially warmed up with the calibration dataset $D$ and are gradually updated with task features during the inference process. 
The label semantic centers remain a true reflection of current conditions by continuously integrating new task features, thereby reducing potential data biases. 
The label semantic center is updated as follows:
\begin{equation}\label{eq21}
    T^c_j = \frac{m_jT^c_j+F_j}{m_j+1}, 
\end{equation}
where $m_j$ denotes the count of tasks, $T^c_j$ is the semantic center and $F_j$ is the current task feature associated with label $j$.

\textbf{Task Separability for Quantization Adjustment.} 
To evaluate correlations between intermediate data and label semantic centers for precise quantization adjustment, we first introduce the concept of similarity degree between task features and label semantic centers.
For assessing similarity, the cosine distance metric \cite{yao2024ferrari} is employed to compare task features with label semantic centers, establishing similarity degrees $\mathbf{T} = \{t_j\},\ j\in\{0,\ldots,n\}$ for each label. 
The similarity degree for label $j$ is formalized as:
\begin{equation}\label{eq22}
    t_j = \xi(F, T^c_j) \in [0,1],
\end{equation}    
where the function $\xi(\cdot)\in [0,1]$ represents the cosine similarity function \cite{nguyen2010cosine}. 
The term $t_j$ quantifies the similarity degree between the task feature $F$ and the label semantic center $T^c_j$, with $t_j=1$ indicating perfect similarity and $t_j=0$ indicating no similarity.
Additionally, to effectively evaluate the quantization precision requirements for task-specific intermediate data, we introduce the concept of task separability, denoted as $S$:
\begin{equation}\label{eq23}
    S = \|\mathbf{T}\|_2\cdot(t_H-t_{SH})\frac{t_H}{t_{SH}},
\end{equation}   
where $t_H$ and $t_{SH}$ represent the highest and second-highest degrees of similarity within the set $\mathbf{T}$, respectively. 
A higher value of $S$ signifies a more pronounced correlation between the intermediate data and the semantic center of the label, suggesting that the inference results are more reliable for the target label. 
This reliability allows for more aggressive quantization, facilitating efficient transmission.


\textbf{Context-Aware Acceleration Strategy.} 
Utilizing the label semantic centers $\mathbf{T_c}$ and task separability $S$, COACH introduces a context-aware acceleration strategy that dynamically adjusts quantization precision and employs an early-exit policy. 
%
The early-exit threshold $S_{ext}$ and quantization precisions thresholds $\mathbf{S}_{adj}$ are initially established using the calibration dataset $D$ to ensure an accuracy loss below 0.5\% \cite{almeida2022dyno}, which only execute one-time prior to running. 
When task separability $S$ exceeds the cache threshold $S_{ext}$, the conditions are met for an early return of inference results. 
The decision-making framework for the early-exit result $R$ is encapsulated by the following equation:
\begin{equation}\label{eq24}
    R = \underset{j \in \{0, \ldots, n\}}{\argmax}\{t_j\in\mathbf{T}\},
\end{equation}   
where the result is ascertained by identifying the highest similarity degree among all labels within $\mathbf{T}$. 
The early-exit policy ensures that decisions are predicated on the most relevant and similar data features, thereby optimizing operational efficiency while maintaining inference accuracy.

In addition, quantization adjustment plays a pivotal role in stabilizing pipeline execution within dynamic network environments.
While offline model partitioning and quantization strategy establishes initial data precision parameters, further minimizing pipeline scheduling bubbles during inference is essential.
This adjustment relies on the premise that greater task separability $S$ enables a more aggressive quantization approach, thus preserving the accuracy of inference results.
By comparing the task separability $S$ with quantization adjustment thresholds $\mathbf{S}_{adj}$, the quantization precision requirement $Q_r$ for the current task is precisely determined.
This precision is then applied to make a real-time decision on quantization adjustment $Q_c$ by minimizing the bubble function:
\begin{equation}\label{eq25}
    Q_c = \argmin_{Q_c \geq Q_r} \{|T_t^\prime - \max\{T_e, T_t^\prime, T_c\}|\},
\end{equation}
where $T_t^\prime$ denotes the transmission time with quantization precision $Q_c$ and real-time bandwidth $B$. 
This formula facilitates task-specific quantization adjustments to further minimize pipeline bubbles, enhancing the efficiency of collaborative inference in dynamic network conditions.

\section{Performance Evaluation}\label{sec:evaluation}
\subsection{Experimental Settings and Baselines}
\textbf{System Implementation.} 
The experimental setup includes a high-performance AMAX deep learning workstation as the cloud server, which is equipped with an Intel Xeon Octa-core processor and 4 NVIDIA A6000 GPUs.
To accurately simulate end devices with varying resources, our platform incorporates both a Nvidia Jetson Xavier NX and a Nvidia Jetson TX2. 
The NX model is equipped with a 6-core NVIDIA Carmel ARMv8.2 CPU and a 384-core NVIDIA Volta GPU, complemented by 8GB of RAM. Conversely, the TX2 features a 4-core ARM Cortex-A57 CPU and a 256-core NVIDIA Pascal GPU, also with 8GB of RAM.
Furthermore, the network setup in our experimental platform includes a 5GHz WiFi router, enabling to establish wireless connections and simulate real-world network conditions. 
We maintain a strict accuracy loss threshold of 0.5\%, ensuring consistent inference performance \cite{almeida2022dyno}.







\textbf{Datasets and Models.} 
The datasets and models in our experiments are detailed as follows:

\begin{itemize} 
    \item \textbf{UCF101} \cite{soomro2012ucf101} is a well-known video dataset primarily used for human action recognition research, comprising 13,320 short videos across 101 action categories.
    For our experiments, we select several frames per second (\eg, 20 frames/sec) to construct continuous inference tasks, arranged chronologically to simulate the real-time activities observed in the videos.

    \item \textbf{ImageNet-100} \cite{krizhevsky2012imagenet} is a curated subset of the extensive ImageNet database, comprising 100 distinct object categories. 
    To evaluate the nature scenes in mobile computing, we split and shuffle the ImageNet-100 dataset with long-tail distribution. 
    This involves allocating a higher selection ratio for more common categories and a lower ratio for less frequent ones.
\end{itemize}


\begin{table}[h]
    \centering
    \caption{Average Inference Latency (ms) for COACH and baselines.}
    \label{tab:all_latency}
    \begin{tabular}{lcccc}
        \hline
                            & \multicolumn{2}{c}{\textbf{Resnet101}} & \multicolumn{2}{c}{\textbf{VGG16}} \\
                            & \textbf{NX} & \textbf{TX2} & \textbf{NX} & \textbf{TX2} \\
        \hline
        \textit{NS}         & 45.16& 62.67 & 29.52 & 52.73 \\
        \textit{DADS}       & 38.11& 49.34 & 24.38 & 39.50 \\
        \textit{SPINN}      & 22.04& 30.10 & 14.03 & 19.97 \\
        \textit{JPS}        & 20.78& 28.31 & 12.52 & 18.13 \\
        \textit{COACH}      & \textbf{15.63}& \textbf{19.11} & \textbf{9.71}  & \textbf{13.37} \\
        \hline 
    \end{tabular}
    \vspace{-3mm}
\end{table}

Our evaluations implement two widely-used DNN models, which achieve high accuracy for real-world applications.

\begin{itemize}

    \item \textbf{VGG16} \cite{simonyan2014very} is characterized by the utilization of $3\times3$ convolutional filters, allowing it to extract intricate features from images for tasks such as image classification and object recognition.
    It consists of 16 processing layers and has 121 million parameters, enhancing its robust performance in diverse applications.

    \item \textbf{ResNet101} \cite{he2016deep} employs Residual Network architecture, which allows gradients to flow through a shortcut path during backpropagation, effectively addressing the vanishing gradient problem in deep networks.
    It features a deep network of 101 layers and incorporates 45 million parameters with DAG topology.
\end{itemize}

\textbf{Baselines and Metrics.} To comprehensively evaluate the inference performance, we compare COACH with the following four baselines:

\begin{itemize}

    \item \textit{Neurosurgeon (NS)} \cite{kang2017neurosurgeon} optimizes the inference latency of individual tasks through the partition strategy of DNNs, specifically tailored for chain topology models.

    \item \textit{DADS} \cite{hu2019dynamic} introduces a model partition strategy for pipeline execution designed to optimize performance in both lightly loaded and heavily loaded networks.


    \item \textit{SPINN} \cite{laskaridis2020spinn} utilizes a dynamic partition strategy and fixed quantization compression, along with an early-exit mechanism to minimize latency under resource constraints. 

    \item \textit{JPS} \cite{duan2023optimizing} proposes a layer-level scheduling algorithm to achieve near-optimal pipeline scheduling for end device computation and transmission stages.

\end{itemize}

To conduct a comprehensive evaluation of COACH and baselines, we employ the following three performance metrics:

\begin{itemize}
    \item \text{Inference Latency (ms)} measures the time required to process each task, critical for assessing the responsiveness of collaborative inference systems.

    \item \text{Transmission Cost (Kb)} evaluates the average transmission overhead for each task, providing insights into the efficiency of data transfer.

    \item \text{System Throughput (it/s)} represents the number of tasks the system can handle per second, providing a concrete measure of the processing capability and operational efficiency of a collaborative inference system. 

\end{itemize}







\subsection{Overall Performance}\label{sec:pfmc}

Our evaluations on latency and throughput are conducted across both high-performance (NX) and low-performance (TX2) devices, across network conditions ranging from 2Mbps to 100Mbps.
We utilize the ImageNet-100 dataset to examine the performance of ResNet101 and VGG16. 
\begin{table}[t]
    \centering
    \caption{COACH performance with context-aware acceleration across different data correlation levels.}
    \label{tab:all_cache}
    \begin{tabular}{lcccccc}
        \hline
                                & \multicolumn{3}{c}{\textbf{Resnet101}}        & \multicolumn{3}{c}{\textbf{VGG16}}           \\
                                & \textbf{Exit.} & \textbf{Ltc.} & \textbf{Trans.} & \textbf{Exit.} & \textbf{Ltc.} & \textbf{Trans.} \\
        \hline
        \textit{NoAdjust}        & -       & 13.63 & 121.0 & -       & 12.02  & 98.0 \\
        \textit{Low}            & 11.86\% & 12.19 & 102.9 & 19.77\% & 10.64  & 75.3 \\
        \textit{Medium}         & 37.41\% & 9.68  & 80.2  & 52.69\% & 7.71   & 44.1 \\
        \textit{High}           & 65.38\% & 6.13  & 40.3  & 76.58\% & 5.77   & 20.6 \\
        \hline
    \end{tabular}
    \vspace{-3mm}
\end{table}

\textbf{Latency Reduction of COACH.} 
To assess the enhanced inference performance of COACH, we record the average inference latency for COACH and baselines. 
Table~\ref{tab:all_latency} illustrates that COACH consistently outperforms four baselines across all inference settings, achieving latency reduction from 22.48\% to 73.59\%. 
Notably, the latency reduction benefit of COACH is more pronounced in scenarios involving low-performance devices (TX2) and complex DNN models (ResNet101), indicating that COACH is effective in optimizing the complex environments of collaborative inference.
Compared to {NS}, {DADS}, and {SPINN}, COACH demonstrates considerable inference improvements, with latency reductions of 69.17\%, 61.26\%, and 36.32\% respectively. 
Moreover, despite the efficient performance of JPS, COACH still manages a 31.75\% latency reduction, showcasing the effectiveness of its quantization strategy and context-aware acceleration.
This performance is especially notable in reducing pipeline bubbles and enhancing computational resource utilization. 
The results underline the ability of COACH to deliver substantial latency improvements, affirming its effectiveness in facilitating lower latency.


\textbf{Context-Aware Acceleration of COACH.} \label{sec:cor}
Our analysis utilizes the UCF101 video dataset to evaluate the performance of context-aware acceleration in COACH across various data correlation levels, ranging from low (random frames), through medium (continuous frames from random videos), to high (continuous frames from sequential videos). 
The NoAdjust scenario, which does not employ context-aware acceleration, serves as a baseline for comparison.
The results reveal a distinct correlation between data correlation levels and performance improvements, particularly quantified by early-exit ratio (Exit.), latency in ms (Ltc.), and transmission costs in Kb (Trans.), as presented in Table~\ref{tab:all_cache}. 
Specifically, the transition from low to high data correlation levels results in early-exit ratio rising from 11.86\% to 65.38\%, with a significant reduction in latency (from 12.19 ms to 6.13 ms) and transmission costs (from 102.9 Kb to 40.3 Kb).
Clearly, COACH is consistently effective at enhancing performance across various data correlation levels, underscoring its adaptability and potential across diverse scenarios. 
Compared to the NoAdjust scenario, incorporating context-aware acceleration within COACH leads to substantial performance enhancements. 
Notably, at high data correlation levels using ResNet101, there is an impressive 65.38\% increase in the early-exit ratio, alongside a significant 66.70\% reduction in average transmission costs and a 55.03\% decrease in latency. 
These improvements are facilitated by the implementation of an adaptive quantization method to optimize the inference process.
Similar benefits are observed with VGG16, which demonstrates the extensive advantages of the early-exit policy and quantization adjustment implemented by COACH under conditions of high data correlation.
These enhancements underline the efficacy of the context-aware acceleration strategy in optimizing inference processes, particularly by further minimizing pipeline bubbles in the inference process and reducing unnecessary data transmissions. 

\begin{figure}[t]
    \centering
    \begin{subfigure}[b]{0.238\textwidth}
        \includegraphics[width=\textwidth]{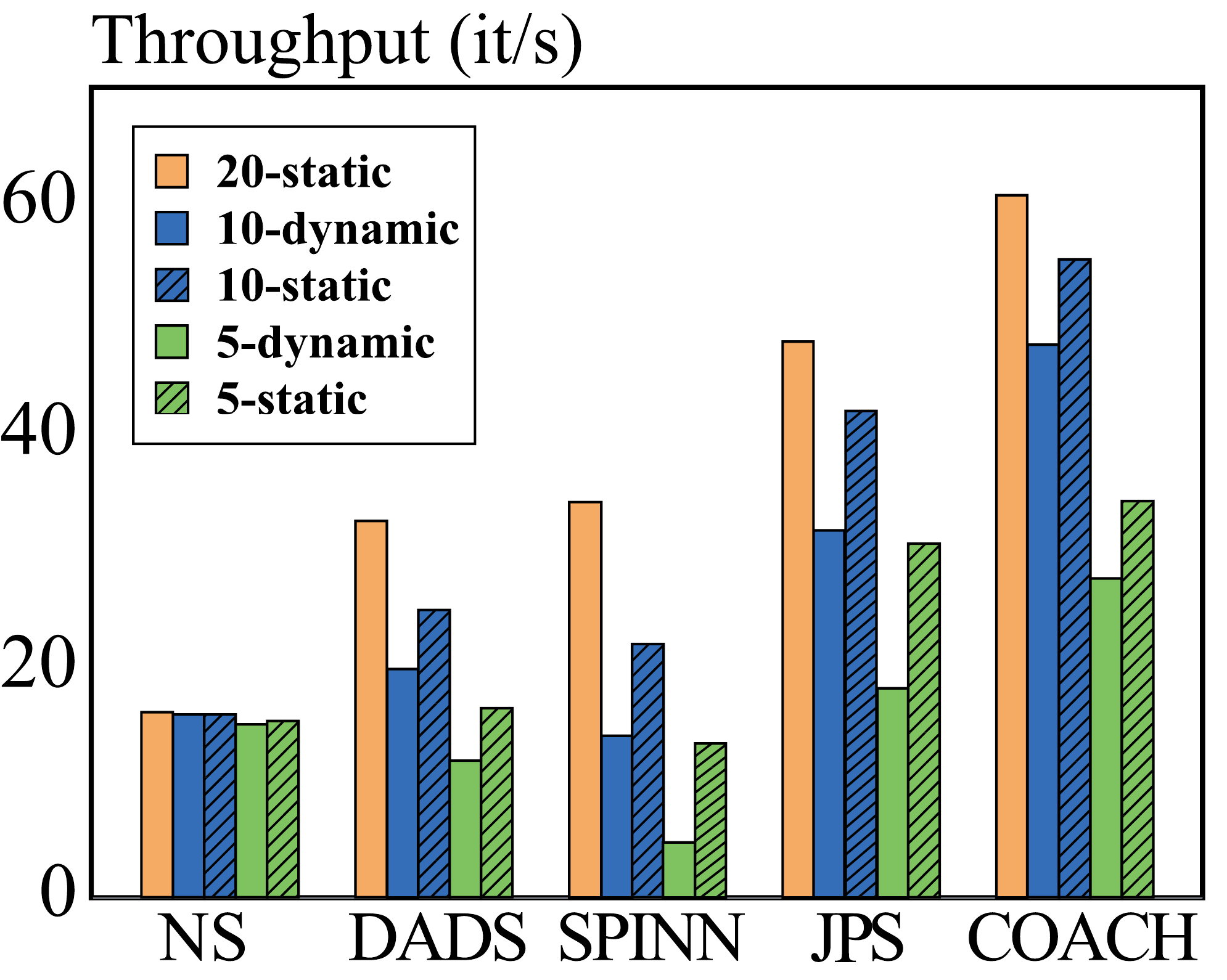}
        \caption{Initial at 20Mbps, reduced to 10Mbps and 5Mbps}
        \label{fig:dy_20}
    \end{subfigure}
    \hfill 
    \begin{subfigure}[b]{0.243\textwidth}
    \includegraphics[width=\textwidth]{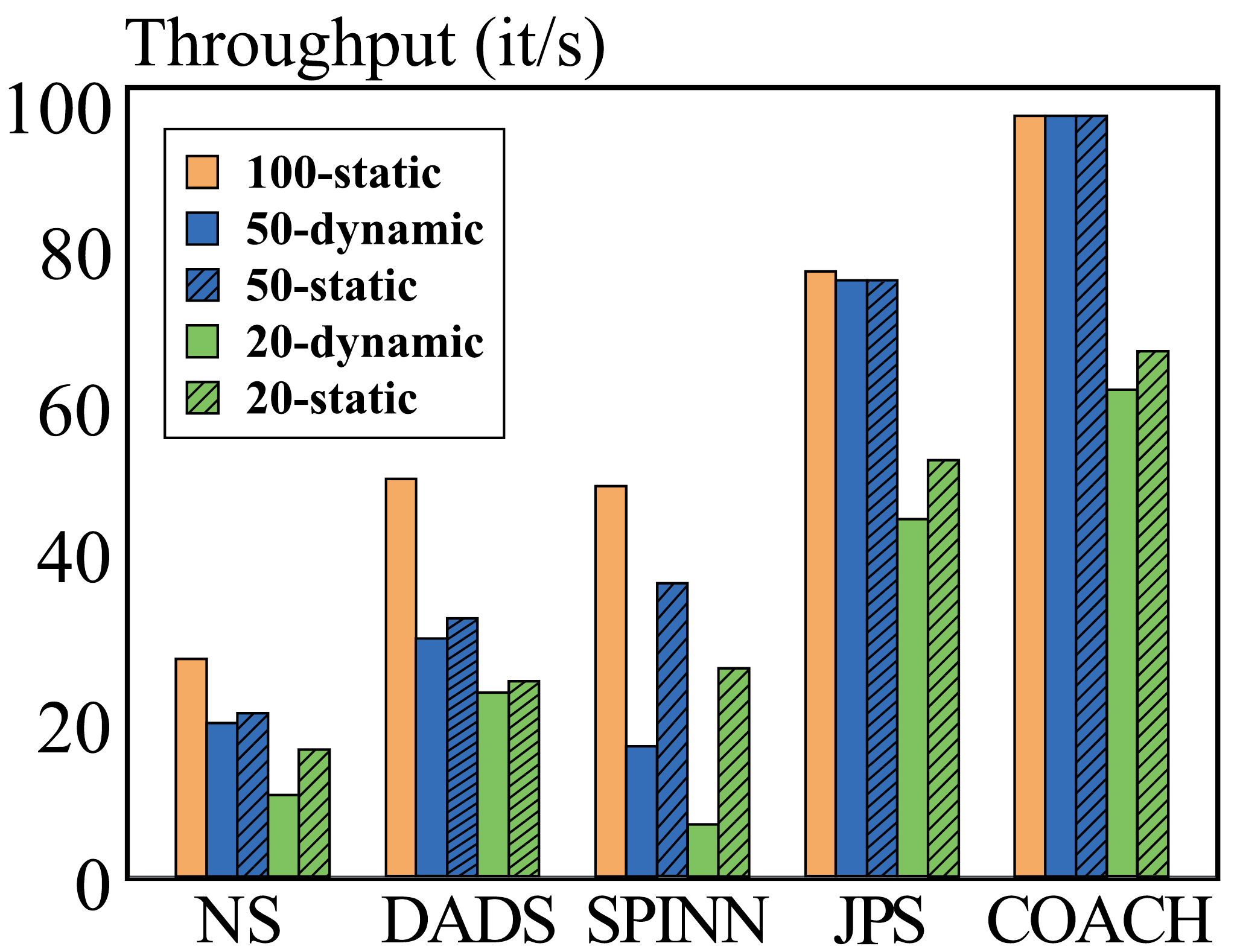}
        \caption{Initial at 100Mbps, reduced to 50Mbps and 20Mbps}
        \label{fig:dy_100}
    \end{subfigure}
    \vspace{-5mm}
    \caption{Adaptability of COACH and baselines in dynamic network conditions.}
    \label{fig:dynamic}
    \vspace{-4.5mm}
\end{figure}

\subsection{System Adaptability in Dynamic Networks}

We assess the adaptability of COACH to dynamic network conditions through a series of experiments conducted on the ImageNet-100 dataset. 
These experiments allow for a detailed analysis of performance across network fluctuations, from initial bandwidths of 20Mbps decreasing to 10Mbps and further to 5Mbps in Fig.~\ref{fig:dynamic}(\subref{fig:dy_20}), as well as from initial bandwidths of 100Mbps decreasing to 50Mbps and 20Mbps in Fig.~\ref{fig:dynamic}(\subref{fig:dy_100}).
Such conditions are known to potentially introduce numerous bubbles in pipeline execution.
We define \textit{static throughput} as the optimal throughput and \textit{dynamic throughput} as the decreased throughput when bandwidth is reduced.
COACH consistently outperforms all baselines with network fluctuations, showcasing enhanced throughput across all bandwidth settings.
In Fig.~\ref{fig:dynamic}(\subref{fig:dy_20}), COACH initially surpasses the throughput of the leading JPS by 1.3$\times$. 
When the bandwidth is reduced to 10Mbps, this advantage expands to 1.4$\times$ compared to JPS, with only a 12\% reduction from its \textit{static throughput} in this bandwidth setting. 
At a further reduced bandwidth of 5Mbps, the throughput of COACH still stands 1.6$\times$ higher than that of JPS, marking just a 15\% decrease from the \textit{static throughput}. 
These results underscore the robustness of COACH, maintaining high throughput consistency and minimizing pipeline bubbles.
In Fig.~\ref{fig:dynamic}(\subref{fig:dy_100}), COACH maintains high throughput consistency, recording 95 it/s even when the bandwidth is halved to 50Mbps, staying 1.2$\times$ higher than JPS. 
This impressive performance showcases COACH can effectively mitigate the impacts of network fluctuations and reduce pipeline bubbles, highlighting its potential and broad prospects for real-world application.

\begin{figure*}[t]
    \centering
    \newcounter{tempfigcnt}\setcounter{tempfigcnt}{\value{figure}}
    \begin{subfigure}{0.55\textwidth}
        \includegraphics[width=\textwidth]{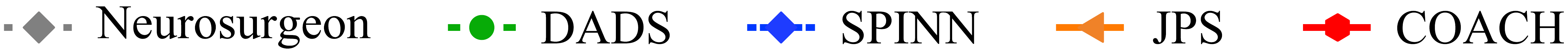}
    \end{subfigure}
    \vspace{-4mm}
    \addtocounter{figure}{-1}\setcounter{figure}{\value{tempfigcnt}}
\end{figure*}
\begin{figure*}[t]
    \centering
    \begin{subfigure}[b]{0.24\textwidth}
        \includegraphics[width=0.9\textwidth]{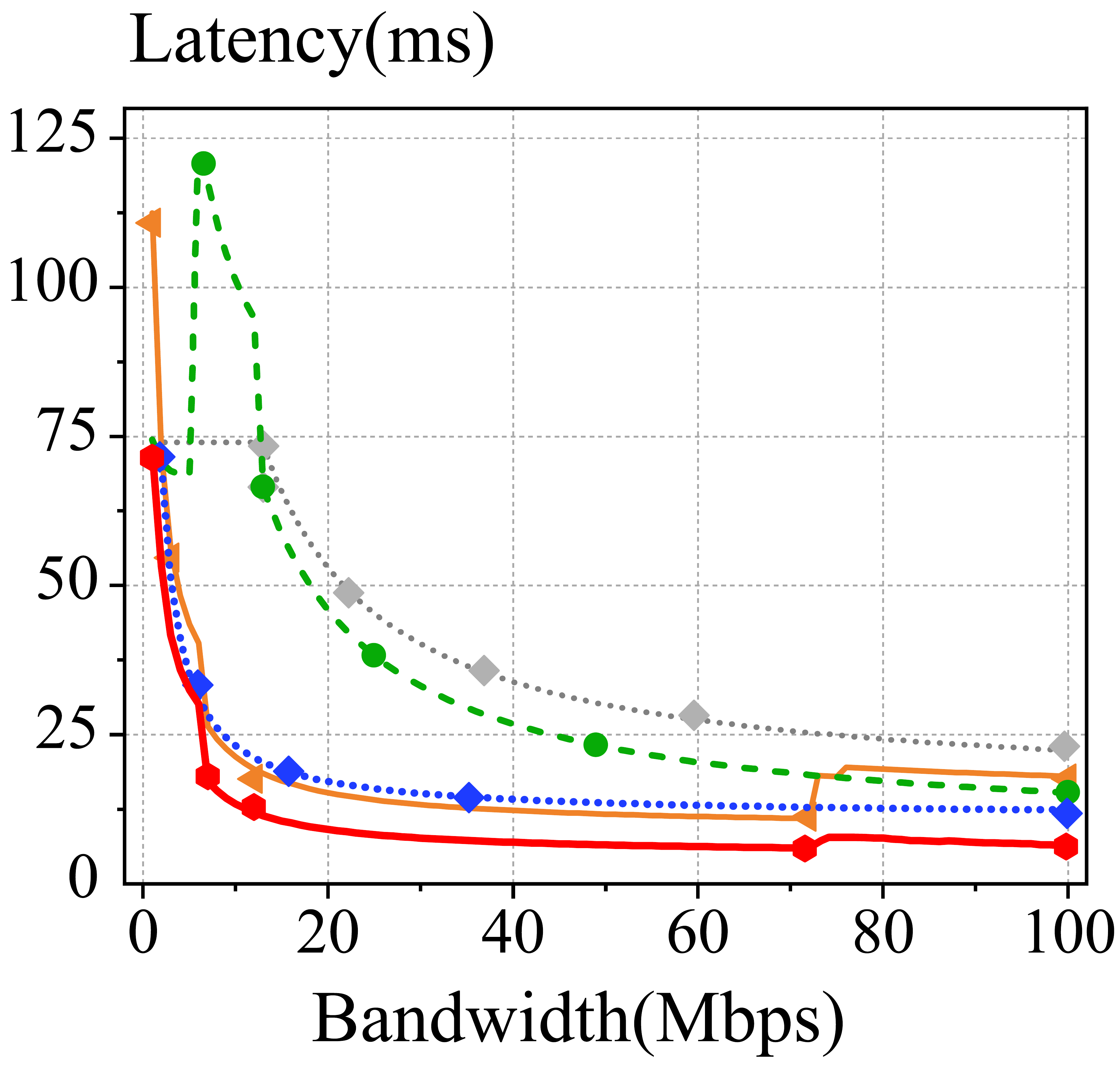}
        \vspace{-2mm}
        \caption{ResNet101 with NX}
        \label{fig:r_10_ltc}
    \end{subfigure}
    \begin{subfigure}[b]{0.24\textwidth}
        \includegraphics[width=0.9\textwidth]{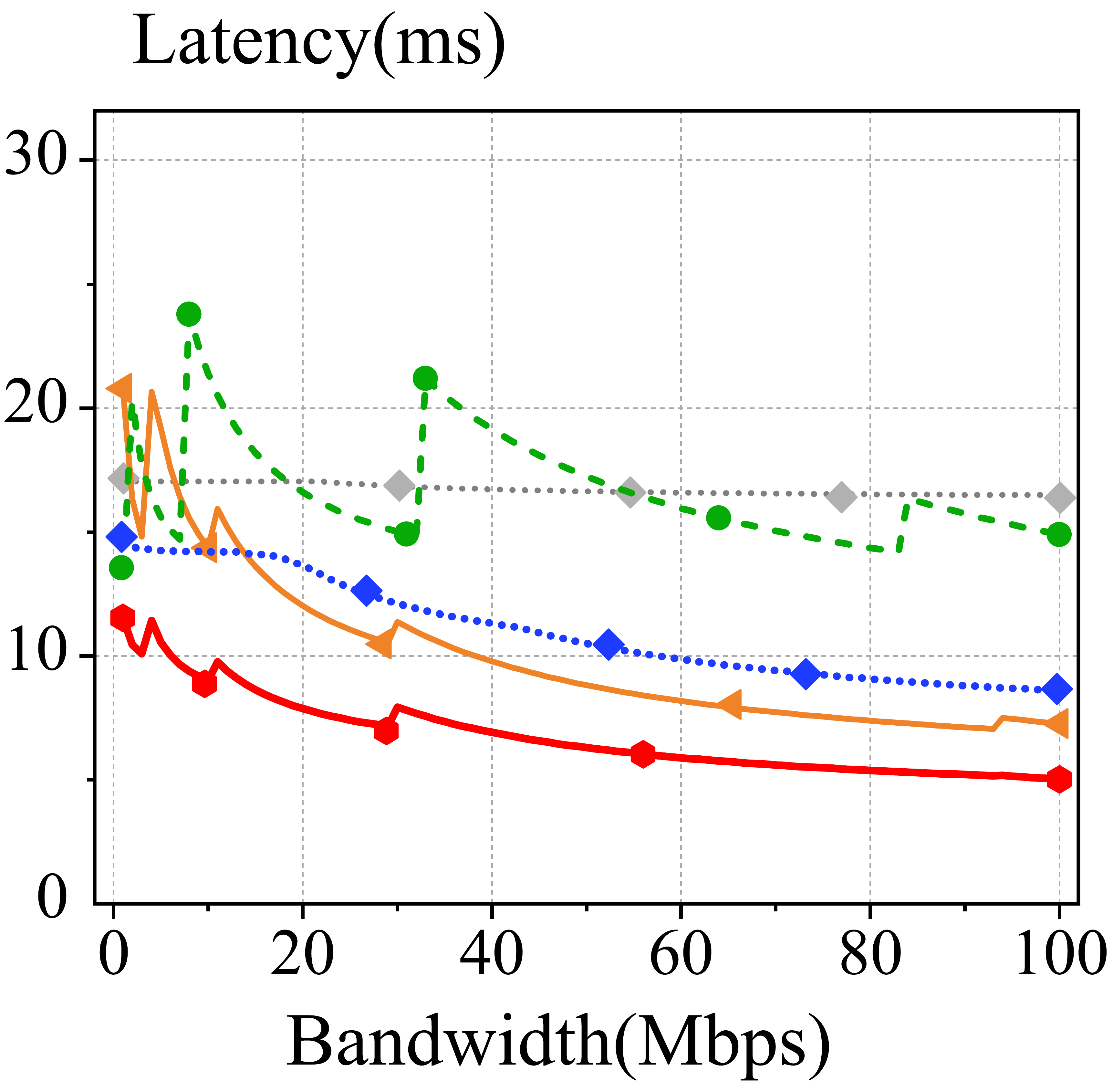} 
        \vspace{-2mm}
        \caption{VGG16 with NX}
        \label{fig:v_10_ltc}
    \end{subfigure}
    \begin{subfigure}[b]{0.24\textwidth}
        \includegraphics[width=0.9\textwidth]{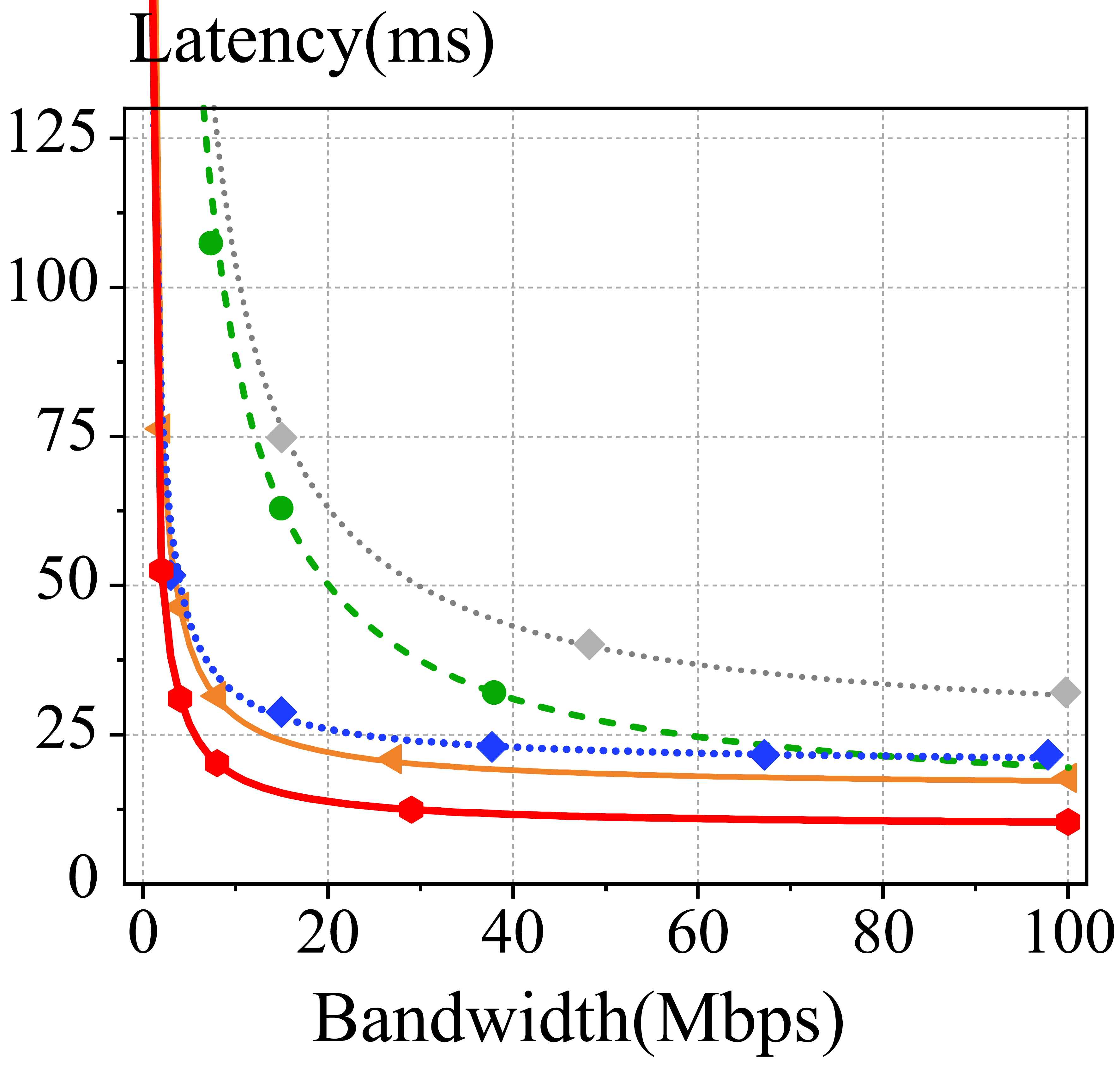}
        \vspace{-2mm}
        \caption{ResNet101 with TX2}
        \label{fig:r_50_ltc}
    \end{subfigure}
    \begin{subfigure}[b]{0.24\textwidth}
        \includegraphics[width=0.9\textwidth]{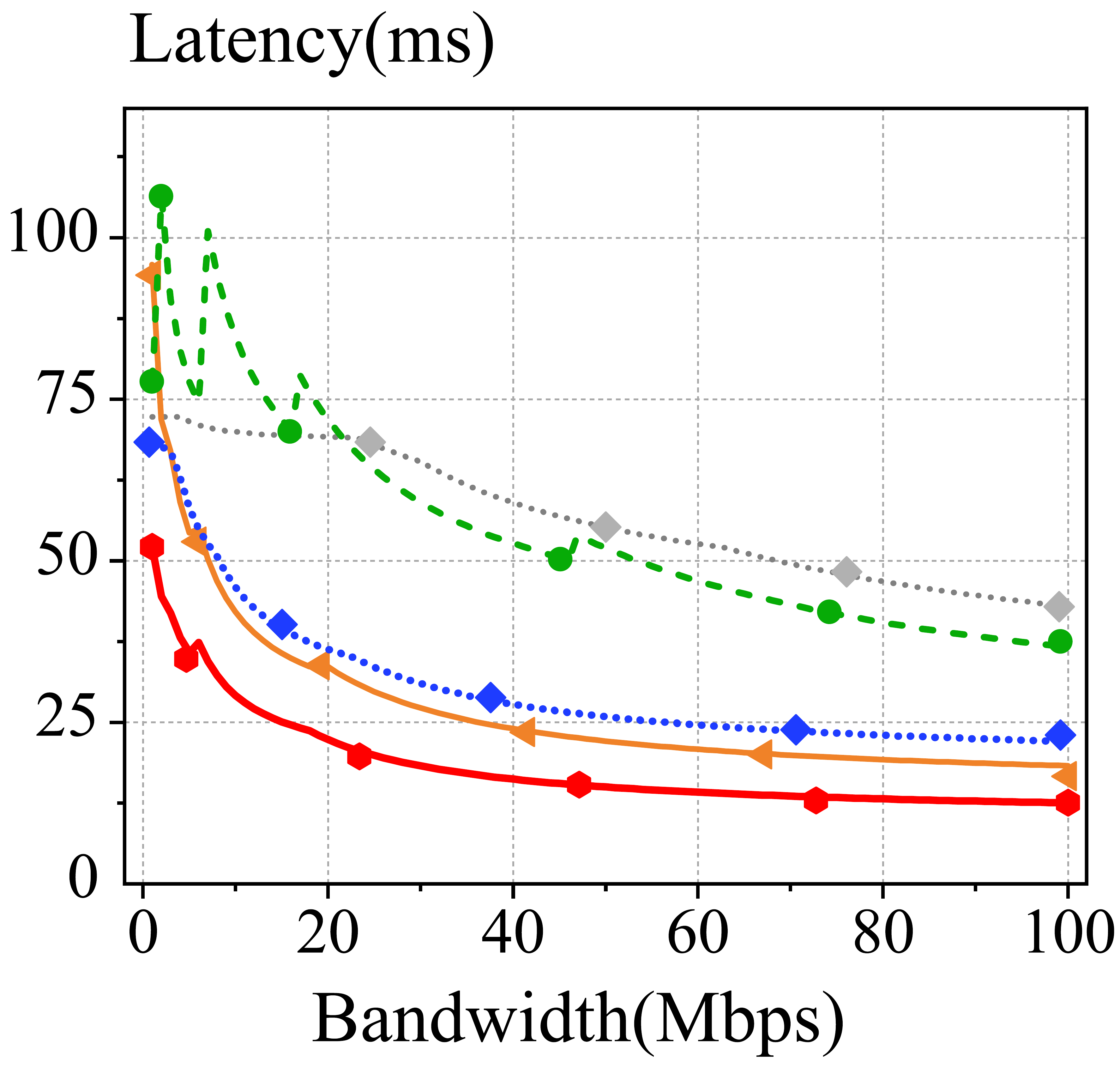}
        \vspace{-2mm}
        \caption{VGG16 with TX2}
        \label{fig:v_50_ltc}
    \end{subfigure}
    \vspace{-1.5mm}
    \caption{Latency of COACH and baselines on Resnet101 and VGG16 in different scenarios.}
    \label{fig:ltc}
    \vspace{-3mm}
\end{figure*}

\begin{figure*}[t]
    \centering
    \setcounter{tempfigcnt}{\value{figure}}
    \begin{subfigure}{0.55\textwidth}
        \includegraphics[width=\textwidth]{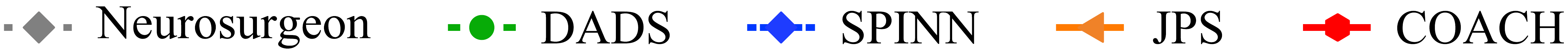}
    \end{subfigure}
    \vspace{-2mm}
    \addtocounter{figure}{-1}\setcounter{figure}{\value{tempfigcnt}}
\end{figure*}

\begin{figure*}[t]
    \centering
    \begin{subfigure}[b]{0.24\textwidth}
        \includegraphics[width=0.9\textwidth]{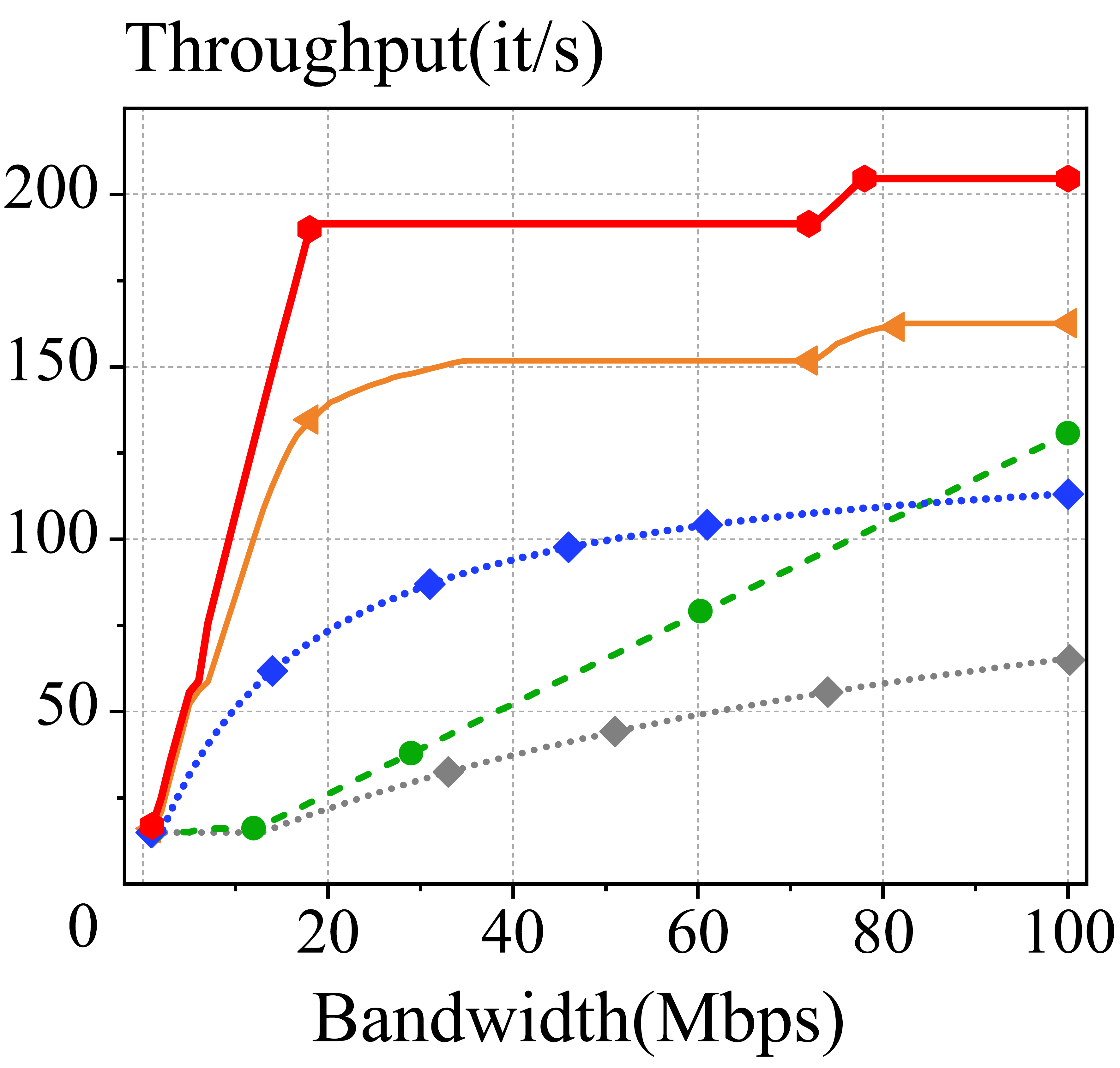}
        \vspace{-2mm}
        \caption{ResNet101 with NX}
        \label{fig:r_10_tp}
    \end{subfigure}
    \begin{subfigure}[b]{0.24\textwidth}
        \includegraphics[width=0.9\textwidth]{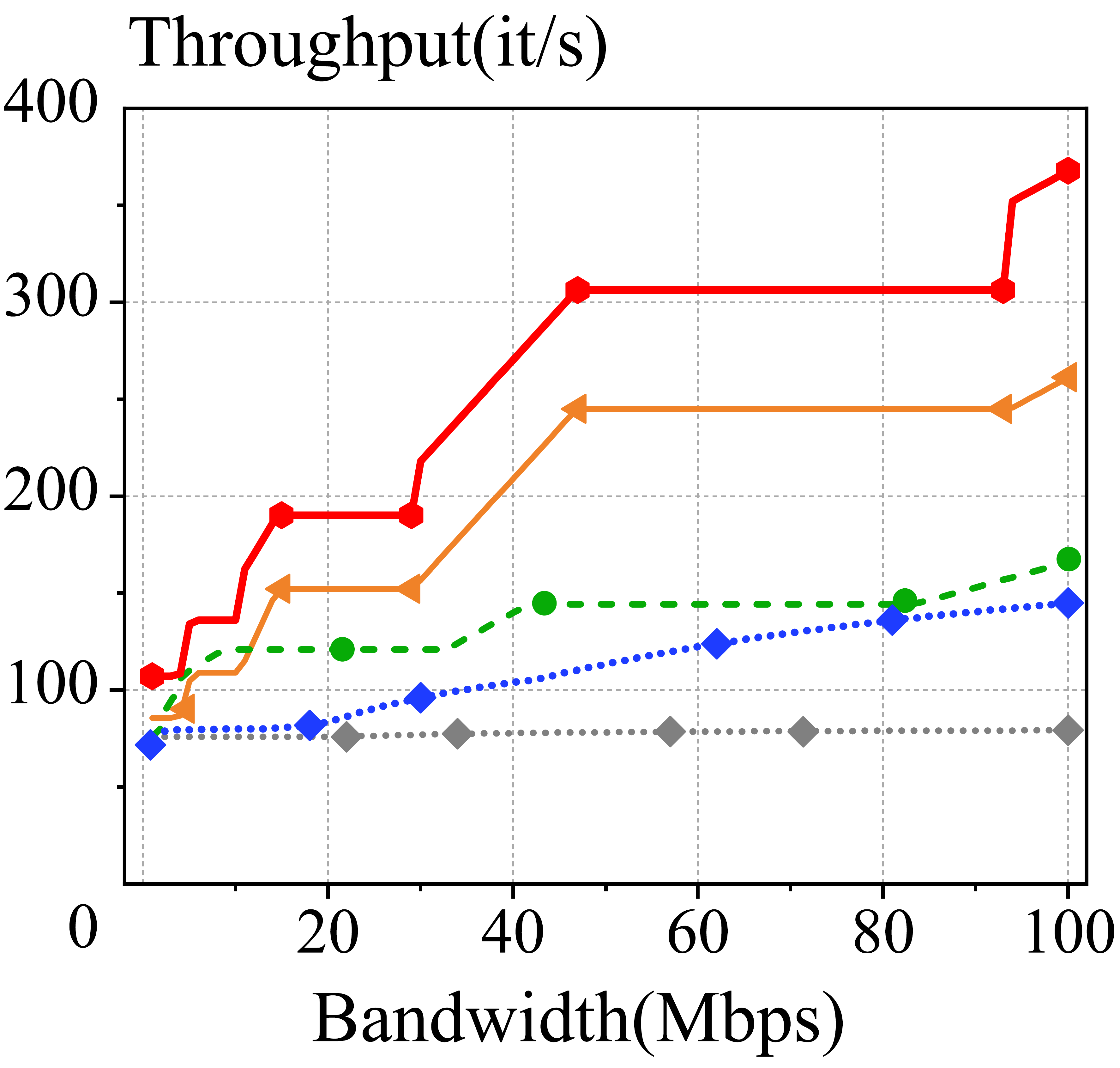} 
        \vspace{-2mm}
        \caption{VGG16 with NX}
        \label{fig:v_10_tp}
    \end{subfigure}
    \begin{subfigure}[b]{0.24\textwidth}
        \includegraphics[width=0.9\textwidth]{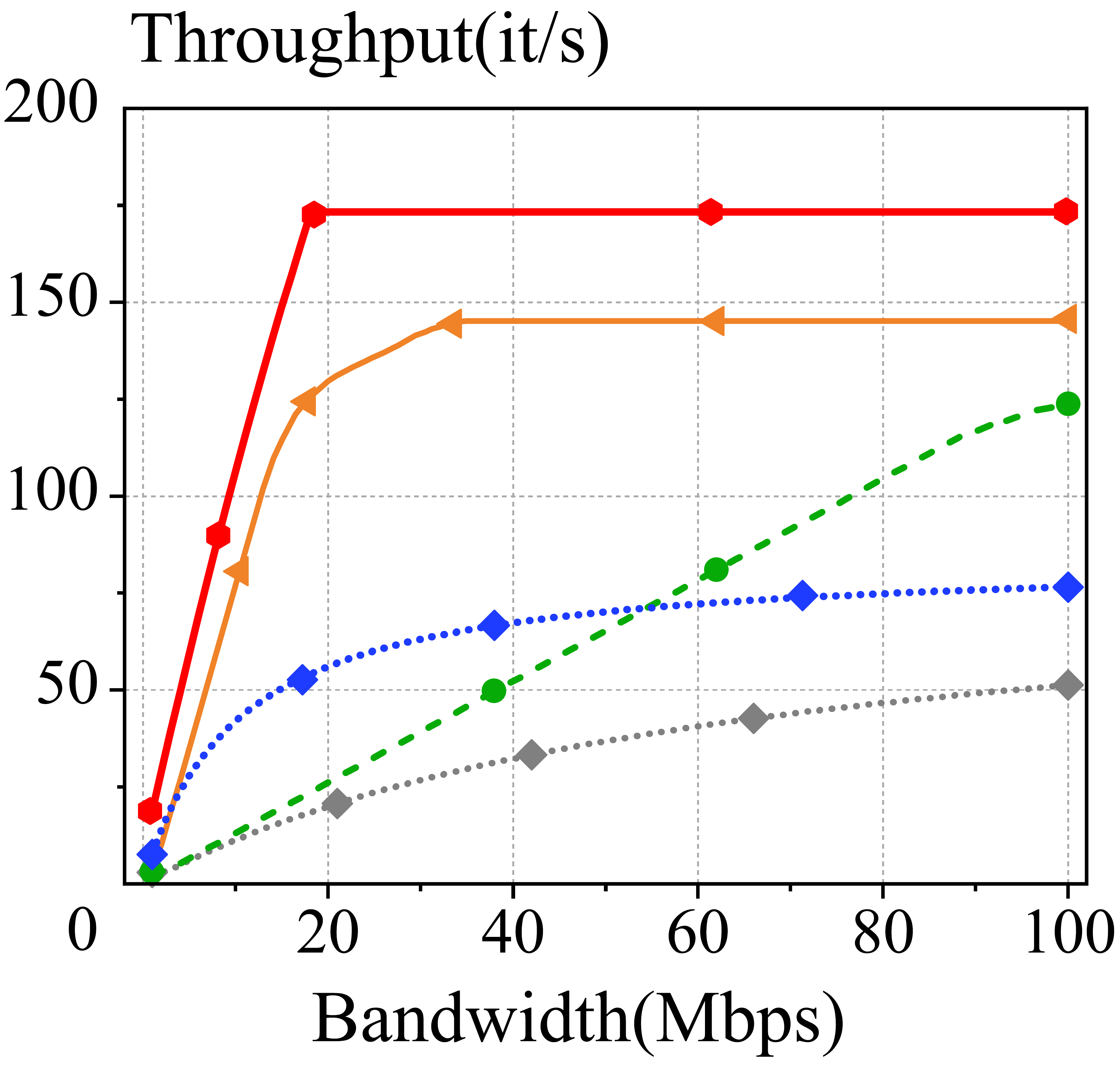}
        \vspace{-2mm}
        \caption{ResNet101 with TX2}
        \label{fig:r_50_tp}
    \end{subfigure}
    \begin{subfigure}[b]{0.24\textwidth}
        \includegraphics[width=0.9\textwidth]{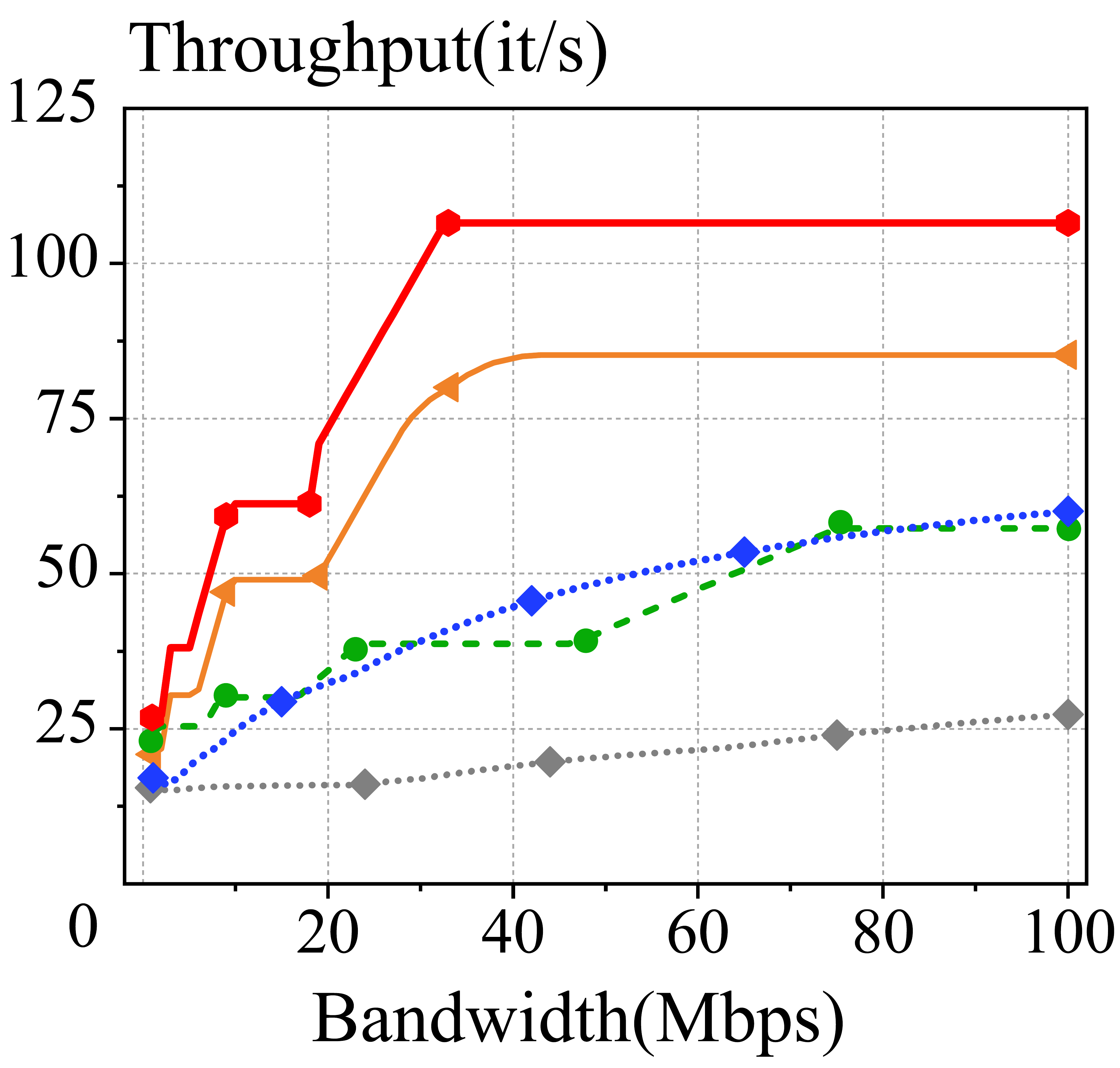}
        \vspace{-2mm}
        \caption{VGG16 with TX2}
        \label{fig:v_50_tp}
    \end{subfigure}
    \vspace{-1.5mm}
    \caption{Throughput of COACH and baselines on Resnet101 and VGG16 in different scenarios.}
    \label{fig:tp}
\vspace{-3mm}
\end{figure*}

\subsection{Impact of Bandwidth on Collaborative Inference}

We conduct a comprehensive evaluation of COACH across diverse settings, leveraging the UCF101 dataset to simulate various network bandwidth scenarios ranging from 1Mbps to 100Mbps. 


\textbf{Latency Performance.} 
The experimental results on collaborative inference latency, as illustrated in Fig.~\ref{fig:ltc} for ResNet101 and VGG16, demonstrate the superiority of COACH in achieving the lowest latency across various bandwidth settings, markedly surpassing baselines.
For ResNet101, under low bandwidth conditions, such as 10Mbps in Fig.~\ref{fig:ltc}(\subref{fig:r_10_ltc}), COACH achieves latency reductions of 72\% and 38\% compared to NS and JPS, respectively.  
In higher bandwidth scenarios (\eg, 50Mbps), COACH reduces latency by 71\% against NS and 42\% against JPS.
Similarly, for VGG16 in Fig.~\ref{fig:ltc}(\subref{fig:v_10_ltc}) and Fig.~\ref{fig:ltc}(\subref{fig:v_50_ltc}), COACH outperforms the baselines across all network conditions.
COACH leverages cache acceleration and quantization adjustment to achieve latency improvements of 55\% and 38\% over NS and JPS in low bandwidth settings such as 5Mbps.
In high bandwidth settings such as 70Mbps, COACH reduces latency by 72\% and 40\% compared to NS and JPS, respectively.
This performance is attributed to the effective utilization of partitioning strategy and context-aware acceleration in COACH, which are optimized for varying network conditions.
These results underline the robustness of COACH in adjusting dynamically to network variations, optimizing both inference performance and resource efficiency.

\textbf{Throughput Performance.} 
Fig.~\ref{fig:tp} illustrates the superior system throughput performance of COACH on ResNet101 and VGG16, underscoring its adaptability across a range of computational resources and network conditions.
For ResNet101 in low bandwidth conditions, such as 10Mbps in Fig.~\ref{fig:tp}(\subref{fig:r_10_tp}), where transmission resources are the bottleneck of the inference process, COACH significantly enhances throughput, achieving increases of 6.2$\times$ compared to NS and 1.6$\times$ compared to JPS.
In high bandwidth conditions such as 50Mbps, where computation resources are the bottleneck, COACH improves throughput by 4.5$\times$ over NS and 1.4$\times$ over JPS. 
This superior performance is due to COACH's effective context-aware quantization adjustment, which optimizes resource utilization and minimizes bottlenecks.
For VGG16, as shown in Fig.~\ref{fig:tp}(\subref{fig:v_10_tp}), COACH significantly enhances throughput, achieving a 9.3$\times$ increase over NS, 3.3$\times$ over SPINN, and 1.8$\times$ over JPS. 
These improvements are attributed to the well-optimized cooperation between the offline and online components in COACH, which ensures efficient pipeline scheduling.


\section{Conclusion}\label{sec:discussion}


In this paper, we propose COACH, a novel framework designed to minimize pipeline bubbles, thereby enhancing the efficiency of collaborative inference systems.
We propose an offline component that incorporates an efficient optimization algorithm for enabling more efficient partition and quantization for pipeline execution.
Furthermore, our online component employs a context-aware acceleration strategy to further reduce pipeline bubbles.
Comprehensive experiments corroborate the superior performance of COACH, showing significant improvements in latency and throughput.


\section{Acknowledgement}\label{sec:Acknowledgement}
This article is supported in part by the National Science Foundation of China (NSFC) under Grants 61936015 and 62132019; in part by the Jiangsu Province Science Foundation for Youths (Grant No. BK20230275); in part by the Anhui Province Science Foundation for Youths (Grant No. 2408085QF185); in part by USTC Research Funds of the Double First-Class Initiative (Grants No. WK2150110033 and No. WK2150110030).

\bibliographystyle{IEEEtran}
\bibliography{ref}

\end{document}